# Analytical modelling for the frequency behavior of two distinct modes in nano-constriction spin Hall nano-oscillator

H. Ghanatian[1,2], M. Rajabali[3], R. Khymyn[4], A. Kumar[4,5,6], V. H. González[4], H. Farkhani[2], J. Åkerman[4,5,6], and F. Moradi[2]

*[1] Department of Mechanical and Electrical Engineering, High Frequency and Digital Electronics Section, University of Southern Denmark, Odense, Denmark*

*[2]ICE-LAB, Electrical and Computer Engineering Department, Aarhus University, Aarhus 8200, Denmark*

*[3] NanoOsc AB, Kista, Sweden*

*[4] Physics Department, University of Gothenburg, 412 96 Gothenburg, Sweden*

*[5]Center for Science and Innovation in Spintronics, Tohoku University, 2-1-1 Katahira, Aoba-ku, Sendai 980-8577 Japan*

*[6]Research Institute of Electrical Communication, Tohoku University, 2-1-1 Katahira, Aoba-ku, Sendai 980-8577 Japan*

Abstract—Nano-constriction spin-Hall nano-oscillators (NC-SHNOs) have garnered considerable interest due to their potential use as efficient and adjustable nano-sized sources of microwave signals, with high-frequency tunability, adaptable design layout, and CMOS compatibility. In order to facilitate system- and circuit-level designs based on the NC-SHNOs, it is essential to have an analytical model capable of predicting the behavior of the NC-SHNO. In this paper, we introduce an analytical model to describe the frequency behavior of a single NC-SHNO in an in-plane magnetic field while considering the Oersted field. The model is divided into two regions based on the direct current value: the "linear-like" and "bullet" modes. Each region is characterized by distinct concepts and equations. The first region, the "linear-like mode," emerges from the nano-constriction edges and progresses toward the center of the active area of the NC-SHNO. In contrast, the second regime, the localized "bullet mode," exhibits negative nonlinearity, where increasing the current will lead to a decrease in frequency. The model's validity is confirmed through experimental data obtained from electrical RF measurements on a single 180nm wide NC-SHNO, and the model demonstrates excellent agreement with experimental data.

Index Terms— Spin Hall nano-oscillator, SHNO, spin Hall effect, analytical model, linear-like mode, bullet mode, high-frequency auto-oscillation, microwave frequency, communication

## I. INTRODUCTION

Spintronics has evolved into a prospective candidate to contribute to numerous applications, encompassing a spectrum from well-established commercial uses, like spin-based non-volatile memories [Zeinali 2018, Bagheriye 2017], to cutting-edge applications such as picoTesla-range magnetic sensors for precise magnetic field detection logic implementations and analog to digital converter [Ghanatian 2023]. Furthermore, spintronic nano-oscillators [Chen 2016] are promising emerging devices with vast potential for applications such as wireless communication [Choi 2014], microwave signal detection [Litvinenko 2020], and neuromorphic computing [Houshang 2022, Torrejon 2017, Ghanatian, 2021, Farkhani 2019, Farkhani 2020]. Among them, spin Hall nano-oscillators (SHNOs) revealed a significant potential as an energy-efficient candidate, specifically in brain-inspired computing, where these nascent nano-oscillators can mimic neurons in a neural network [Moradi 2019, Zahedinejad 2020].

These nonlinear oscillators, introduced by Demidov et al. [Demidov 2014], are bow-tie-shaped nano-constrictions (NCs), which are mainly in the form of heavy metal / ferromagnetic (HM/FM) bilayer. They benefit from the spin Hall effect, where the pure spin current is produced by passing direct current (DC) through an HM. Due to high spin-orbit interaction in the HM, the spins with opposite directions accumulate at opposite metal interfaces, diffuse, and exert torque on the magnetization of the adjacent FM. When the torque is strong enough to overcome the FM intrinsic damping, it will lead to a steady state precession of magnetization, so-called auto-oscillation. Through the anisotropic magnetoresistance effect (AMR), this can then be transformed into a microwave voltage. The frequency of the auto-oscillation can be tuned by various parameters, such as the current density in the active area of NC-SHNO (i.e., the center of the constriction) and the external magnetic field strength and angle.

The main advantages of NC-SHNOs are their straightforward fabrication, flexibility in material choices, frequency tunability, and direct optical access to the active auto-oscillating region. They also offer low power consumption, compact size, and compatibility with existing semiconductor fabrication processes, turning them into a promising candidate to contribute to future high-density, high-speed, and energy-efficient brain-inspired neural networks [Zahedinejad 2022, Kumar 2023].

The analytical modeling of NC-SHNOs plays a crucial role in understanding their behavior, optimizing their performance, and performing hybrid system design (e.g., integrating NC-SHNO and CMOS circuits) for different applications such as neuromorphic computing implementation. Analytical modeling enables the prediction and analysis of essential characteristics of NC-SHNOs, such as their oscillation frequencies. Hence, it can offer valuable insights into the fundamental mechanisms governing NC-SHNO behavior, which can be utilized to optimize the design, stability, and efficiency. Unlike the spin torque nano-oscillators (STNOs) [Boone 2009, Chen, T 2015], there has been no solid study of an analytical model for NC-SHNO. Albertsson et al. have studied a compact macro spin-based model developed for in-plane (IP) magnetized three-terminal MTJ-SHNOs [Albertsson 2019]. Still, no further studies have been provided for a stand-alone NC-SHNO to predict its characteristics. Here, an analytical model of the oscillation frequency of a single 180 nm wide NC-SHNO is presented. This model, in which the Oersted field is considered, predicts the NC-SHNO frequency tunability in an IP external field at 60 mT. The analytical model is divided into two parts depending on the device operation mode, namely linear-like and bullet modes. In linear mode, increasing the current passing through the device ($I_{SHNO}$) results in an increase in frequency (positive nonlinear frequency shift). However, in bullet mode, the relationship is the opposite (negative nonlinear frequency shift). The presented model in linear-like mode is based on [Slavin 2009] considering the SHNO structure (bilayer structure and bow-tie-shaped nano-constriction) and taking into the Oersted field and positive nonlinearity. In the bullet mode, an analytical model for the amplitude of a spin wave is proposed based on [Slavin 2005] by considering the device structure, the Oersted field, and a negative nonlinear shift. The model is implemented in Verilog-A, which is important for device, circuit- and system-level simulations. It is important to be able to initially model a single NC-SHNO oscillation frequency as a single spintronic device and predict all evolved modes before implementing it at the circuit- and system levels. The results were validated by the experimental data obtained through the electrical RF measurements. This paper is organized as follows. In Section II, the fabrication process and electrical measurements are elaborated. The modeling of NC-SHNO is presented in Section III, and finally, Section V concludes this paper.

## II. Experimental Section

The details for the fabrication process and the electrical measurements are described in sections A and B, respectively, and section C is dedicated to demonstrating the extracted results and parameters from the electrical measurements.

### A. SHNO fabrication process

The NC-SHNOs were fabricated on high-resistance silicon substrates (ρ > 10,000 μohm-cm) bearing thin layers of W (5 nm) / Py (5 nm) ($Ni_{80}Fe_{20}$) / $Al_2O_3$ (4 nm) heterostructures [Behera 2022]. The stack layers were deposited using an AJA Orion 8 magnetron sputtering system with a base pressure lower than $3\times10^{-8}$ Torr. Using a combination of e-beam lithography (Raith EBPG 5200) and Ar-ion beam etching (Oxford Dry etch 400 Plus), the resulting stacks were patterned into 8×12 $\mu m^2$ rectangular mesas with bow-tie-shaped NC. The fabrication details can be found in [Kumar 2022]. The coplanar waveguide (CPW) was then defined using a laser writer-based lithography followed by deposition and liftoff of Cu (800 nm) / Pt (20 nm) bilayer.

Fig. 1(a) shows the final NC-SHNO connected to the rest of the setup for a steady-state auto-oscillation (AO) measurement at room temperature to determine the behavior of NC-SHNOs. The inset depicts the Ground-Signal-Ground (GSG) CPW encompassing a single 180 nm wide NC-SHNO, while Fig. 1(b) is the top-view scanning electron microscopy (SEM) image of the same size NC-SHNO.

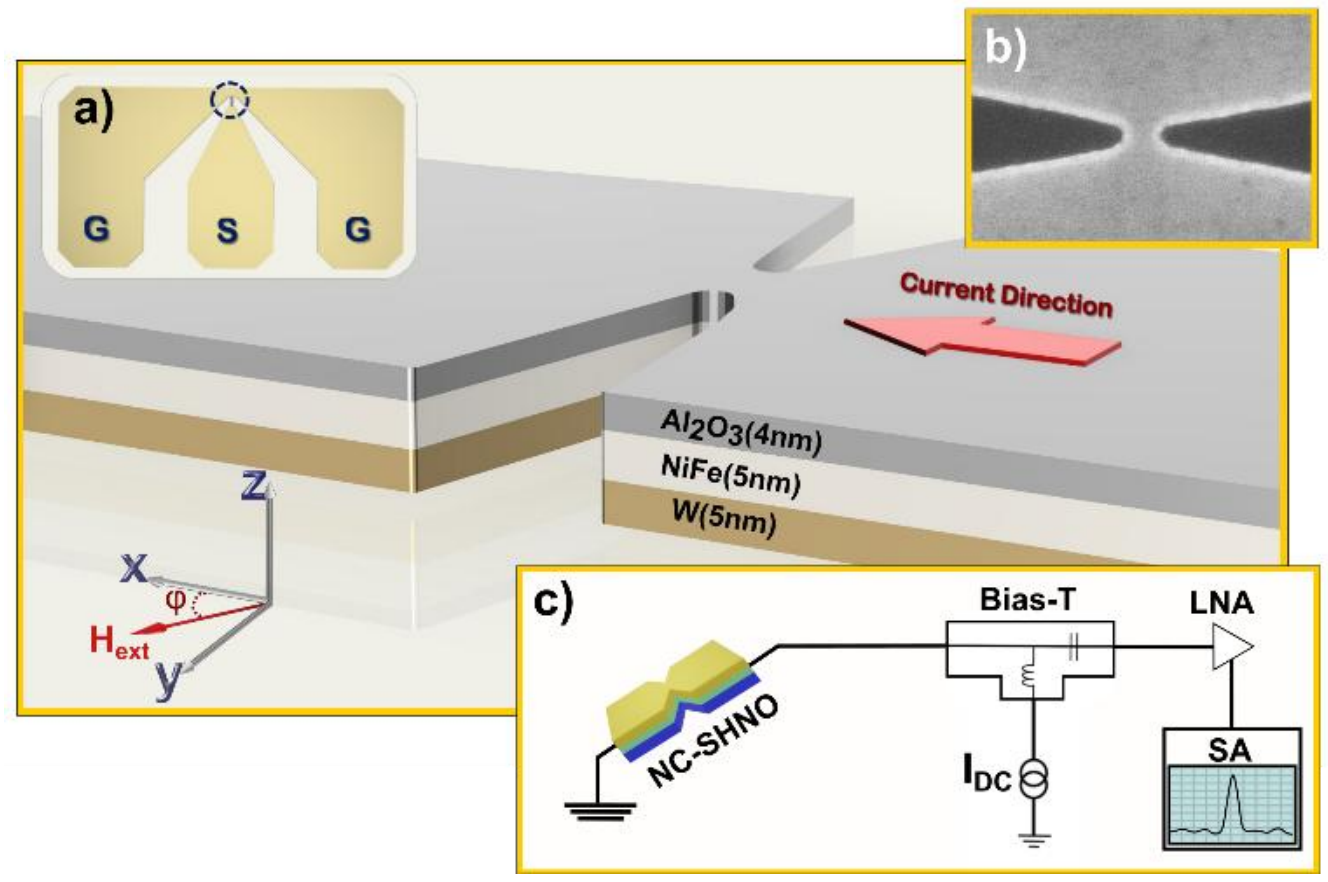

Figure 1 : (a) Device schematic and the current and field direction in combination with the device layer structure. Inset: the Ground-Signal-Ground pads connecting to the SHNO. (b) SEM image of the 180 nm wide NC-SHNO. (c) Schematic representation of the auto-oscillation measurement setup.

### B. Electrical measurements

The microwave measurements are performed using an experimental setup depicted in the scheme in Fig. 1(c), initialized by passing direct current (DC) along the device while an external magnetic field is applied at a certain IP angle $\varphi$ [Fig. 1(a)]. The applied DC, which flows to the device through the DC port of a bias-tee, generates a spin current in the HM, which diffuses into the adjacent FM layer and leads to a sustainable precession of the FM magnetization at a frequency in the GHz range. The generated microwave signal then goes through the high-frequency port of bias-T and eventually to a low noise amplifier (LNA) before being captured by a Rhode & Schwarz FSV-40 spectrum analyzer (SA). Based on signal linewidth, a resolution bandwidth (RBW) of 1 MHz and a video bandwidth (VBW) of 1 kHz have been chosen.

### C. Results and discussions

The obtained power spectral density (PSD) from SA is presented in Fig. 2(a). The current sweep measurement is performed at a fixed IP angle of 22 deg at an external field of 60 mT while scanning SHNO current ($I_{SHNO}$) from 1 to 2.2 mA.

Before analyzing the captured spectrum, it is crucial to eliminate the background noise level that spans the entire measured frequency range. However, the background noise level is not evenly distributed throughout the scanned frequency range. Therefore, the spectra at very low DC (10 μA) are obtained in the first measurement trace to determine the noise floor for the entire frequency range. This noise floor is subtracted from the captured PSD spectra to produce the AO spectra. The AO spectra are represented in dB over noise, as seen in Fig. 2(a). Furthermore, the measured PSDs are fitted to symmetric Lorentzian functions to extract the peak frequency, signal coherency (linewidth), and peak power. The extracted parameters are plotted as a function of $I_{SHNO}$ and shown in Fig. 2(b-d).

The extracted data in Fig. 2(b-d) imply that the SHNO signal comprises two distinct regions. The first region, starting from the threshold current (1.15 mA) to 1.47 mA, is a "linear-like mode" [Mazraati 2018].

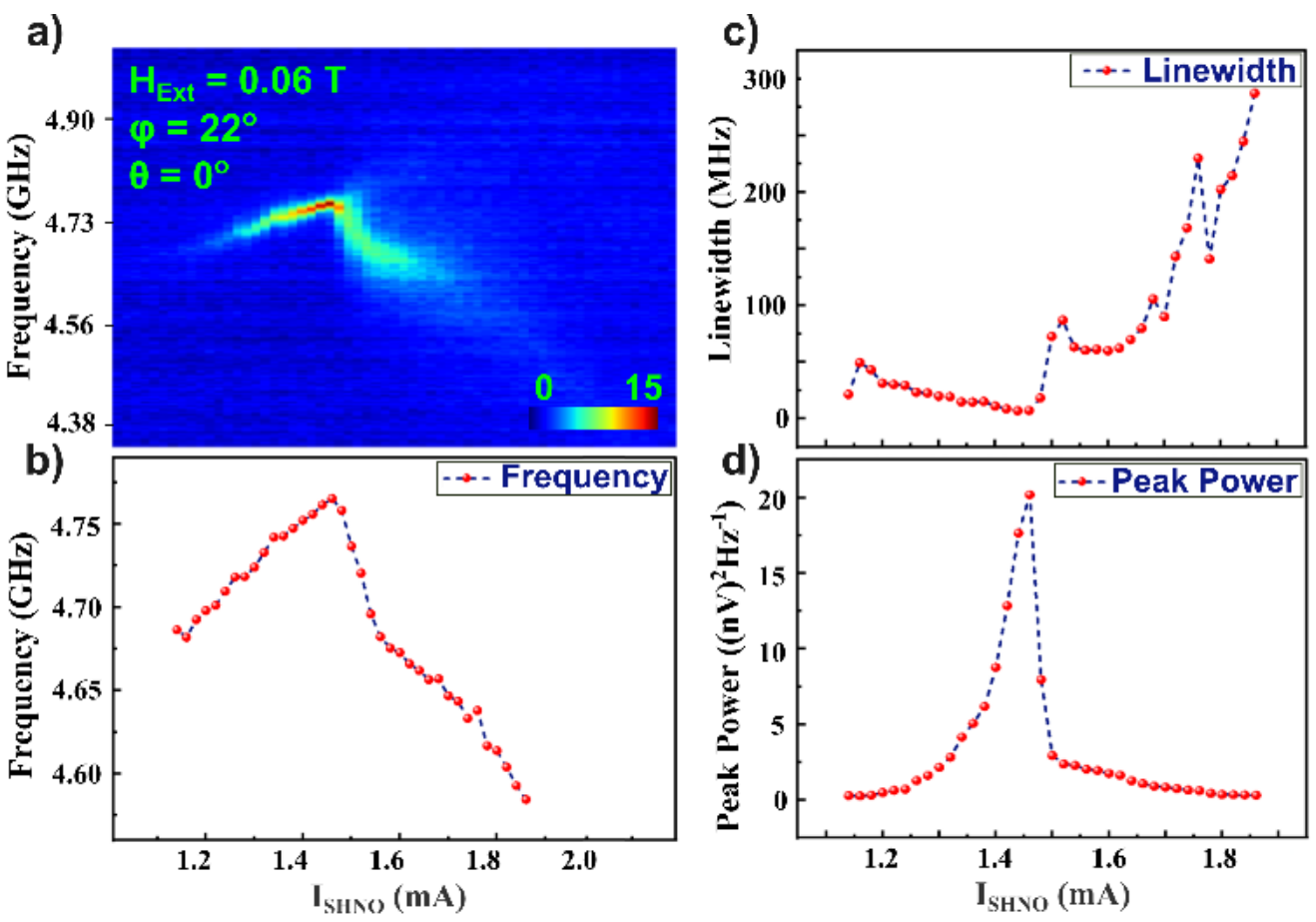

Figure 2: (a) Power spectral density *vs.* $I_{SHNO}$ of the NC-SHNO at a 0.06 T IP field. Extracted (b) peak frequency, (c) linewidth, and (d) peak power from (a) as a function of $I_{SHNO}$.

As $I_{SHNO}$ increases, the evolution of this mode begins from the edges of the nano-constriction and gradually grows towards the center of the active area [Dvornik 2018, Mazraati 2018]. At 1.47 mA, the mode changes. In this second regime, a negative nonlinearity (the frequency decreases with increasing $I_{SHNO}$), a widening linewidth, and a decreasing power are observed, which are the main characteristics of a bullet mode [Dvornik 2018]. Moreover, by pumping more magnons into the system (a consequence of increasing $I_{SHNO}$), magnons start repelling each other after a certain value, eventually leading to mode dissipation [Rajabali 2023].

Since all extracted data, i.e., nonlinearity, linewidth, and peak power, behave differently within these two modes, to provide a realistic model, we need to divide the signal into two distinct zones. Zone 1 is within the current range of (1.15 mA-1.47 mA), and Zone 2 is within (1.47mA-1.7mA), which are attributed to the linear-like and bullet mode, respectively. Each zone is studied separately with its relative concept and equations, which will be discussed in detail in the next section.

## III. Modeling Section

This section presents an analytical model for the NC-SHNO frequency behavior. The magnetization dynamics within macroscopic approximation can be predicted by the Landau-Lifshitz-Gilbert (LLG) equation as follows [Liu 2011]:

$$\frac{\partial \vec{m}}{\partial t} = -\gamma \vec{m} \times \left(\vec{H}_{\mathrm{Oe}} + \vec{H}_{\mathrm{eff}}\right) + \alpha \vec{m} \times \frac{\partial \vec{m}}{\partial t} + \frac{\hbar}{2e\mu_0 t_{\mathrm{FM}} M_s} J_S \left(\vec{m} \times \vec{\sigma}_{\mathrm{SHE}} \times \vec{m}\right). \quad (1)$$

In Eq. (1), $\gamma$, $\alpha$, $M_s$, $\mu_0$ , $\hbar$, $e$, $\eta$, and $t_{FM}$ are the gyromagnetic ratio, the Gilbert damping constant, the saturation magnetization, the vacuum permeability, the reduced Planck constant, the electron charge, the spin Hall angle, and the FM thickness, respectively. $\vec{\sigma}_{\mathrm{SHE}}$ , $\vec{m}$, and $\vec{H}_{\mathrm{eff}}$ are the polarization direction of the pure spin current, the FM magnetization direction, and the effective magnetic field, respectively. $\vec{H}_{\mathrm{Oe}}$ is the Oersted field, and according to Ampère's law, it is considered $J_{SHNO}$ d/2 [Liu 2011], in which $J_{SHNO}$ and d are the current density flowing through the NC-SHNO and HM thickness, respectively. Here, $J_S$ $\hbar$/2e represents the spin current density injected into the FM layer.

The model is based on the nonlinear auto-oscillator theory of microwave signal generation proposed by Slavin and Tiberkevich [Slavin 2009]. In the literature, the frequency behavior of spin-based oscillators is modeled according to the universal model of auto-oscillators with nonlinear frequency shifts and negative damping.

The first step of modelling of NC-SHNO is to calculate $H_{\mathrm{eff}}$ and $\varphi_{\mathrm{eff}}$ by deriving the following equations:

$$H_{\mathrm{eff}} \cos(\theta_{\mathrm{eff}}) \cos(\varphi_{\mathrm{eff}}) = H_{\mathrm{ext}} \cos(\theta_{\mathrm{ext}}) \cos(\varphi_{\mathrm{ext}}) + H_{\mathrm{A}} \cos(\theta_{\mathrm{eff}}) \cos(\varphi_{\mathrm{eff}}) \quad (2a)$$

$$H_{\mathrm{eff}} \cos(\theta_{\mathrm{eff}}) \sin(\varphi_{\mathrm{eff}}) = H_{\mathrm{ext}} \cos(\theta_{\mathrm{ext}}) \sin(\varphi_{\mathrm{ext}}) - H_{\mathrm{Oe}} \quad (2b)$$

$$H_{\mathrm{eff}} \sin(\theta_{\mathrm{eff}}) = H_{\mathrm{ext}} \sin(\theta_{\mathrm{ext}}) - H^{\mathrm{demag}} \sin(\theta_{\mathrm{eff}}) \quad (2c)$$

$H_{\mathrm{ext}}$, $\theta_{\mathrm{ext}}$, and $\varphi_{\mathrm{ext}}$ are the strength, out-of-plane (OOP), and IP angle of the external magnetic field, respectively. In addition, $\theta_{\mathrm{eff}}$, $\varphi_{\mathrm{eff}}$, and $H^{\mathrm{demag}}$ are the OOP and IP angles of the effective magnetic field and the demagnetization field ($4\pi M_S$), respectively. Since the current flows into the NC-SHNO along the x-axis, according to Ampère's law, $\vec{H}_{\mathrm{Oe}}$ is generated along the y-axis.

In this model, the $H_{\mathrm{ext}}$ is IP, which leads to simplifying Eq. (2a) – (2c) [Albertsson 2019].

$$H_{\mathrm{eff}} = H_{\mathrm{ext}} \frac{\cos \varphi_{\mathrm{ext}}}{\cos \varphi_{\mathrm{eff}}} + H_{\mathrm{A}} \quad (3a)$$

$$H_{ext} \frac{\sin \varphi_{eff}}{\cos \varphi_{eff}} \cos\varphi_{ext} + H_A \sin\varphi_{eff} - H_{ext} \cos\varphi_{ext} = -H_{Oe} \quad (3b)$$

where $H_{\mathrm{eff}}$ and $\varphi_{\mathrm{eff}}$ can be calculated from Eq. (3a) and (3b).
The frequency of operation $\omega$ is calculated according to Eq. (4) [Albertsson 2019].

$$\omega \approx \omega_0 + N_{Lin} \bar{p} \quad (4)$$

where $\omega_0$, $N_{Lin}$ and $\bar{p}$ are the ferromagnetic resonance (FMR) frequency, the nonlinear frequency shift coefficient for the linear mode, and the average dimensionless spin-wave power, respectively. $\omega_0$ can be described by the Kittel equation according to Eq. (5). This equation is achieved by assuming only IP fields for an extended thin film, i.e., when $H_{demag} = M_s$ [Slavin 2009]. However, in the NC-SHNO, the edge of the constriction is modifying this demagnetization field, which can be taken into account approximately by replacing $M_s$ with some effective value $M_{Lin}$.

The linear expansion of the frequency on power (Eq. (4)) is valid in the relatively narrow range of the oscillation amplitudes. Increasing the amplitude beyond this range makes the nonlinearity coefficient N dependent on the power itself. This is especially noticeable in the transition from the linear to the bullet mode, where the precession amplitude changes rapidly, together with the nonlinearity sign switching from positive to negative. Thus, it is convenient to define separate coefficients $N_{Bullet}$ and $M_{Bullet}$ for the soliton mode. $M_{Lin}$ and $M_{Bullet}$ are shown in Table. 1.

$$\omega_0 = \gamma \sqrt{\left(H_{\mathrm{eff}} - \sin^2\varphi_{\mathrm{eff}} H_{\mathrm{A}}\right)\left(H_{\mathrm{eff}} + H^{demag}\right)} \quad (5)$$

To calculate $\bar{p}$, Eq. (23) is utilized [Slavin 2009].

$$\bar{p} = \frac{Q\eta}{Q+\zeta}\left[1 + \frac{\exp\left(-\frac{\zeta+Q}{Q^2\eta}\right)}{E_\beta\left(\frac{\zeta+Q}{Q^2\eta}\right)}\right] + \frac{\zeta-1}{\zeta+Q} \quad (6)$$

where $Q$ and $\zeta$ are the nonlinear damping coefficient and the effective noise power, respectively. In this equation $\beta = -(1 + Q)\, \zeta / Q^2 \eta$ and $E_n(x) = \int_1^\infty \frac{e^{-xt}}{t^n} dt$, and $Q$ and $\zeta$ are considered fitting parameters. They will be tuned to fit the experimental data. The supercritical parameter, $\zeta$, is calculated as follows [Liu 2011]:

$$\zeta = \frac{I_{\mathrm{HM}}}{I_{\mathrm{th_HM}}} \quad (7)$$

where $I_{th}$ is the threshold current, which means the minimum required current to generate AO in the NC-SHNO, and $I_{HM}$ is the effective current that passes through the HM. In this structure, the injected current into the bilayer is divided into two parts: one flows through the HM and the rest to the FM layer [King 2014]. In other words, the resistance of NC-SHNO ($R_{SHNO}$)equals $R_{FM} \parallel R_{HM}$, where $R_{FM}$ and $R_{HM}$ are FM and HM resistances, respectively [King 2014]. Hence, $I_{HM}$ can be defined as follows:

$$I_{\mathrm{HM}} = I_{\mathrm{SHNO}} \frac{R_{\mathrm{FM}}}{R_{\mathrm{FM}} + R_{\mathrm{HM}}} \quad (8)$$

Eq. (9) is utilized to calculate $I_{th}$:

$$I_{\mathrm{th}} = \frac{\Gamma_{\mathrm{G}}}{\sigma} \quad (9)$$

where $\Gamma_G$ is the damping term and can be defined from [Eq. (104b) in [Slavin 2009]] and $\sigma$ is obtained from Eq. (10):

$$\sigma = \frac{\theta_{\mathrm{SH}} g \mu_{\mathrm{B}}}{2eA_{\mathrm{HM}} t_{\mathrm{FM}} M^{demag}} \cos(\varphi_{\mathrm{eff}}) \quad (10)$$

where $g$, $\mu_B$, $A_{HM}$, and $\theta_{SH}$ are spectroscopic Lande factor, Bohr magneton, HM cross-sectional area, and spin Hall angle, respectively.

The oscillation frequency of the NC-SHNO in the bullet mode can be modeled according to Eq. (11) [Gerhart 2007]:

$$\omega \approx \omega_0 - |N_{Bullet}||a_0|^2 \quad (11)$$

where $N_{Bullet}$ is the nonlinear frequency shift coefficient in bullet mode. A solitonic type spin wave mode, also known as a "spin wave bullet," can be strongly localized and non-propagating by an IP magnetized free magnetic layer. While it is a widely recognized fact that solitonic wave packets in conservative systems tend to be unstable in two-dimensional space, a spin wave bullet with a particular amplitude ($a_0$) can be stabilized by the interaction between the spin-polarized current, which generates negative damping, and the inherent positive magnetic damping of the free layer. $a_0$ can be calculated from Eq. (12) [Slavin 2005]:

$$\frac{\sigma I}{\Gamma_G} = \frac{\chi}{\eta_2(qa_0) - a_0^2 \eta_4(qa_0)} \quad (12)$$

$$\eta_n(qa_0) \equiv \int_0^\infty f(x/qa_0)\psi^n(x)x dx \quad (13)$$

where $f(x/qa_0)$ is a steplike current distribution which means if x < 1 f(x) = 1 otherwise f(x) = 0. In this work, the dimensionless function, $\psi(x)$, is considered as sech(x) [Slavin 1990], and $\chi$ is 1.86 [Slavin 2005]. q is a parameter that measures the degree of nonlinearity relative to exchange-originated dispersion. By using Eq. (12), Eq. (13), and $\psi(x)$ = sech(x), Eq. (14) can be obtained to calculate $a_0$:

$$\frac{\sigma I}{\Gamma} = \chi \left[ qa_0 \cosh(qa_0) + \ln(\cosh(qa_0)) + qa_0{}^3 \tanh(qa_0) - \frac{a_0{}^3}{6} tanh^3(qa_0) - \frac{1}{3} qa_0{}^3 \tanh^2(qa_0) + \frac{2}{3} a_0{}^2 \ln|sech(qa_0)| \right]^{-1}$$

(14)

The values of the required parameters are presented in Table 1.

Fig. 3 shows the $I_{SHNO}$ *vs.* $a_0$ for q of 1.62, 1.82 and 1.92. The result is obtained from Eq. (14). The plots exhibit a distinct minimum that aligns with the minimum amplitude ($a_0 = a_{th}$) of a bullet generated at the microwave threshold through spin-polarized current. Each current has two amplitudes of $a_0$ ($a_0 < a_{th}$ or $a_0 > a_{th}$). For $a_0 < a_{th}$, the amplitude of the mode will gradually diminish until it reaches the background noise level, while for $a_0 > a_{th}$, the amplitude will rise until it reaches a stable value. Hence, this work considers the amplitudes of $a_0$, which are more than $a_{th}$.

In this model, the value of 1.82 is chosen for "q" as it corresponds to the amplitude of $I_{SHNO}$ where the bullet mode becomes evident. The bullet mode is characterized by a decreasing frequency with increasing $I_{SHNO}$.

Table 1. The values of the required parameters

| Parameter | Value |
|---|---|
| IP external magnetic field ($H_{ext}$) | 0.06 T |
| IP angle of external field ($\varphi_{ext}$) | 22 Deg |
| HM resistivity ($\rho_{HM}$) | 280 m Ω.cm |
| Ferromagnetic resistivity ($\rho_{FM}$) | 40 m Ω.cm |
| Nonlinear damping ($Q$) | 9 |
| Effective noise power ($\eta$) | 0.002 |
| Anisotropy field ($H_A$) | 0.003 T |
| $M_{Lin}$ | 0.52 T |
| $M_{Bullet}$ | 0.64 T |
| $N_{Lin}$ | 11.78 G (rad/s) |
| $N_{Bullet}$ | 20 G (rad/s) |
| Gilbert damping ($\alpha$) | 0.012 |
| Gyromagnetic ratio ($\gamma/2p$) | 29 $GHz/T$ |
| Spin Hall angle ($\theta$) | - 0.5 |
| Constant form factor of the bullet ($\chi$) | 1.86 |
| Nonlinearity factor of the bullet (q) | 1.82 |
| Thickness of heavy metal ($t_{HM}$) | 5 nm |
| Nano-constriction width (w) | 180 nm |
| The length of the mesa ($L_{SHNO}$) | 12 µm |
| The width of the mesa ($w_{SHNO}$) | 8 µm |

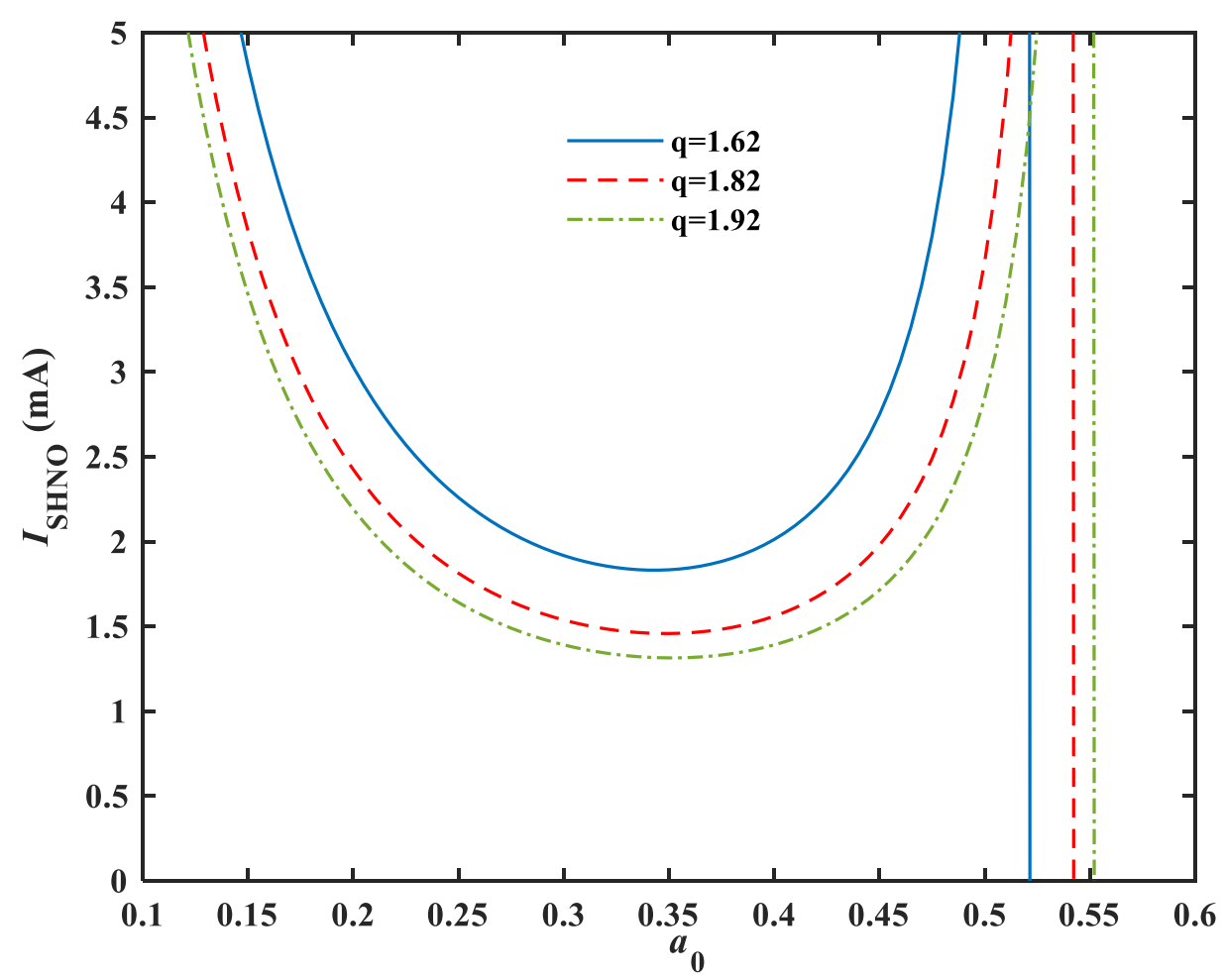

Figure 3: $I_{SHNO}$ *vs.* the amplitude of bullet, $a_0$, for the different nonlinear parameter, q.

Fig. 4 illustrates the oscillation frequency plotted *vs.* $I_{SHNO}$ in different distinct regions: the linear-like and bullet mode. The figure presents a comparison between the model and the experimental measurements, implying that there is an excellent agreement between them. As mentioned, Eq. (4) and Eq. (11) are applied to calculate the frequency in linear-like and bullet modes, respectively.

The gray area represents the transition between the linear-like and bullet modes, a rapid transition from the repulsive interaction between magnons in the linear-like mode to the attractive one, which forms a spin-wave bullet at higher currents.

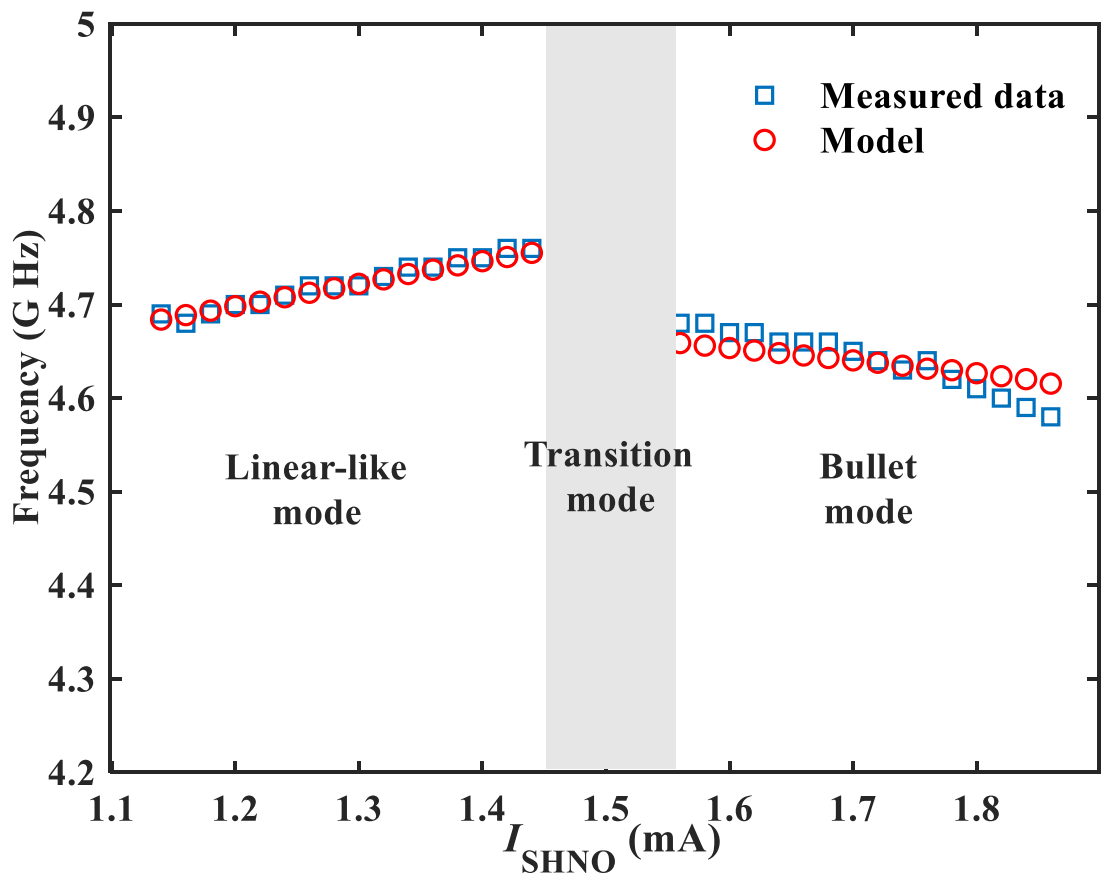

Figure 4: A 180 nm wide NC-SHNO oscillation frequency *vs.* $I_{SHNO}$ in linear-like and bullet modes.

## IV. CONCLUSION

This paper introduces an analytical model to describe the frequency behavior of a single 180 nm wide NC-SHNO while considering the Oersted field. The frequency behavior can be categorized into two distinct regions: linear-like and bullet-mode regions. Our approach to modeling the frequency behavior of NC-SHNOs is based on a generalized framework that incorporates auto-oscillators with nonlinear frequency shifts and negative damping. The implementation of this model demonstrates excellent agreement with experimental data that opens up the possibility for designing and simulating hybrid NC-SHNOs/CMOS circuits, accelerating the integration of these spintronic devices into future electronics.

## ACKNOWLEDGMENT

This research work is supported by the European Union's Horizon 2020 Research and Innovation Program Grant No. 899559 "SpinAge", DOI 10.3030/899559.